\documentclass[aps, twocolumn, pra, showpacs]{revtex4}
\usepackage{graphicx}
\usepackage{amsmath}
\usepackage{amsfonts}
\usepackage{amssymb}
\usepackage{epsfig}

\begin{document}

\title{Fidelity susceptibility and quantum adiabatic condition in
thermodynamic limits}
\author{Shi-Jian Gu}
\email{sjgu@phy.cuhk.edu.hk}
\affiliation{Department of Physics and ITP, The Chinese University of Hong Kong, Hong
Kong, China}

\begin{abstract}
In this work, we examine the validity of quantum adiabatic theorem in
thermodynamic systems. For a $d$-dimensional quantum many-body system, we
show that the duration time $\tau_0$ required by its ground-state adiabatic
process does not depend on the microscopic details, but the scaling
dimension of the fidelity susceptibility $d_a$. Our result, therefore,
provides a quantitative time scale of the quantum adiabatic theorem in
thermodynamic systems. The quantum adiabatic theorem might be violated in
case that the scaling dimension of the fidelity susceptibility is larger
than the system's real dimension ($d_a>d$).
\end{abstract}

\date{\today }
\maketitle




Quantum phase transition (QPT)\cite{Sachdev} is one of the most active
research fields in condensed matter physics. For a quantum many-body system
described by a Hamiltonian $H(\lambda )$, a QPT occurs as its ground-state
property undergoes a significant change at a transition point $\lambda _{c}$%
. In order to study QPTs, people usually work on the lowest eigenstate of $%
H(\lambda )$. In practice, if there is no other mechanisms to change the
lowest eigenstate, but drive the system from one phase to the another by
changing the driving parameter $\lambda $ directly, one should ensure the
validity of the quantum adiabatic theorem.

The quantum adiabatic theorem states that a quantum state will not transit
to the system's other states of different eigenenergy if the driving
Hamiltonian changes slowly enough in time. The theorem is an extremely
intuitive concept because its validity relies on the criterion of the
\textquotedblleft slowness". This criterion, for an arbitrary $D$-level
system, has been improved step by step in the last several decades \cite%
{LDLandau,CZener32,Schiff,MGellMann51,MVBerry84,EFarhi2001,JLiu,DMTong2005,DMTong2007}.
However, the relation between the \textquotedblleft slowness" and thermodynamic
properties, such as dimensionality and various critical
exponents etc, have been paid few attention. Therefore, for a $d$%
-dimensional quantum thermodynamic system, how to define \textquotedblleft
slowness" or its relation to statistical quantities remains a fundamentally
important question. To answer this question in a quantitative way is the key
motivation of the present work.

In this report, we start from the time-dependent Schr\"{o}dinger equation,
and show that the leading transition probability from the ground state to
excited states at the perturbation level is proportional to the fidelity
susceptibility \cite{WLYou07,PZanardi0701061}, which was proposed recently
in the fidelity studies on the QPTs \cite%
{HTQuan2006,Zanardi06,HQZhou0701,LCVenuti07,SChen07,MFYang07,SJGu072,HMKwok07,SYang08FS,XWang012105,SJGu08,JZhang,KWSun,SJGUreview}%
. Then we are able to use the scaling dimension of the fidelity
susceptibility $d_{a}$ (called quantum adiabatic dimension hereafter \cite%
{SJGu08}) to quantify the scale of the duration time required by the quantum
adiabatic theorem. A general inequality is established for the slowness
criterion (in terms of the duration time) in the thermodynamic limit.

We take the linear quench process, in which the driving Hamiltonian is
turned on linearly with the time $t$, as an example. In this case, the
duration time $\tau _{0}$ for sufficient slowness should satisfy $\tau
_{0}\gg \kappa L^{d_{a}}$ where $\kappa $ is independent of $L$. Therefore,
if we require that a physically acceptable duration time is proportional to
the system size, which is about the order of the Avogadro constant ($%
6.02\times 10^{23}$) for a realistic system, then the two limits of $%
N(=L^{d})\rightarrow \infty $\ and $\tau _{0}(\propto N)\rightarrow \infty $%
\ do not commute with each other in case that the quantum adiabatic
dimension $d_{a}>d$ (in other words $\infty \neq \infty ^{\mu }$ for $\mu
\neq 1$), hence the quantum adiabatic theorem might break down. We finally
examine the validity of the quantum adiabatic theorem in a few of many-body
systems, including the one-dimensional transverse-field Ising model \cite%
{Sachdev}, the Lipkin-Meshkov-Glick(LMG) model \cite{LMGmodel}, and the Kitaev
honeycomb model \cite{Kitaev}. In these models, $d_{a}>d$ at their
corresponding critical point, hence the quantum adiabatic theorem is violated
around the critical point. In the Kitaev honeycomb model, moreover, the quantum
adiabatic dimension in the gapless phase is 2+ln (for a size dependence of
$N^{2}\ln N$), which is still larger than the real dimension 2, so the
adiabatic theorem might break down in the whole phase.

To begin with, we consider a general $d$-dimensional quantum many-body
system of length $L$ and size $N=L^{d}$. Its Hamiltonian reads
\begin{equation}
H(\lambda )=H_{0}+\lambda H_{I},  \label{eq:Hamiltonian0}
\end{equation}%
where $H_{I}$ is the driving Hamiltonian, $\lambda =\lambda _{i}+t/\tau _{0}$
denotes its time-dependent strength with $\tau _{0}$ being the duration time
scale and $\lambda _{i}$ the starting point. To be consistent with the
time-dependent perturbation theory, we let $\{|\phi _{n}(t)\rangle \}$
define the complete set of eigenstates of the instant Hamiltonian $H(t)$,
i.e. $H(t)|\phi _{n}(t)\rangle =\epsilon _{n}(t)|\phi _{n}(t)\rangle $.
According to the quantum adiabatic theorem, the ground state of the system
is always $|\phi _{0}(t)\rangle $ if the driving Hamiltonian $H_{I}$ is
turned on slowly enough (here we exclude those cases of the ground-state
level-crossing). Then we can always use the adiabatic ground state $|\phi
_{0}(t)\rangle $ to study QPTs in the parameter space of $\lambda $.

According to quantum mechanics, the system's state can be expressed as a
linear combination of the adiabatic eigenstates,
\begin{equation}
|\Psi (t)\rangle =\sum_{n}a_{n}(t)|\phi _{n}(t)\rangle ,  \label{eq:GenWF}
\end{equation}%
which is required to satisfy the time-dependent Schr\"{o}dinger equation,
i.e.
\begin{equation}
i\frac{\partial }{\partial t}|\Psi (t)\rangle =H(t)|\Psi (t)\rangle .
\label{eq:SchrodingerEq}
\end{equation}%
Here we set $\hbar =1$. Introducing the unitary transformation%
\begin{equation}
a_{n}(t)=\tilde{a}_{n}(t)\exp \left( -i\int^{t}\epsilon _{n}t^{\prime
}dt^{\prime }\right) ,
\end{equation}
and combining Eq. (\ref{eq:GenWF}) and Eq. (\ref{eq:SchrodingerEq})
together, we can obtain
\begin{eqnarray}
\frac{\partial \tilde{a}_{m}}{\partial t} &=&-\tilde{a}_{m}\langle \phi
_{m}|\partial _{t}\phi _{m}\rangle  \label{eq:fsmgd} \\
&&-\sum_{n\neq m}\frac{\langle \phi _{m}|\partial _{t}H|\phi _{n}\rangle
\tilde{a}_{n}}{\omega _{nm}}\exp \left( -i\int^{t}\omega _{nm}dt^{\prime
}\right) ,  \notag
\end{eqnarray}%
where $\partial _{x}\phi \equiv \partial \phi /\partial x$ and $\omega
_{nm}\equiv \epsilon _{n}-\epsilon _{m}$. The quantum adiabatic theorem is
based on the approximation that if the second term on the right hand side of
the above equation is small enough compared with the first term, then the $m$%
th state will keep its position except for an accumulation of a Berry phase
from the first term of Berry connection. Such an approximation holds true
for a finite-size system with a finite-size energy gap, but should be
treated very carefully for a thermodynamic system in which the gap might
vanish and long-range correlations appear in various distinct ways.

Now we suppose the Hamiltonian evolves from $\lambda $ to $\lambda +\delta
\lambda $ during a finite time interval $\Delta t$, that is $\Delta t=\delta
\lambda \tau _{0}$. The $\delta \lambda $ is small enough for the validity
of the time-dependent perturbation theory. We will return to this
requirement later. At time $t=0$, $\tilde{a}_{0}=1,\tilde{a}_{m}=0$, so the
system is at the ground state $\phi _{0}(t=0)$, then at $t=\Delta t$, we
have, to the first order,%
\begin{eqnarray}
\tilde{a}_{0} &\simeq &1-\frac{1}{\tau _{0}}\int_{0}^{\Delta t}\langle \phi
_{0}|\partial _{\lambda }\phi _{0}\rangle dt,  \label{eq:berrycon} \\
\tilde{a}_{m} &\simeq &-\frac{1}{\tau _{0}}\int_{0}^{\Delta t}\frac{%
H_{I}^{m0}}{\omega _{0m}}\exp \left( -i\int^{t}\omega _{0m}dt^{\prime
}\right) dt,  \label{eq:hopingcon}
\end{eqnarray}%
where $H_{I}^{nm}=\langle \phi _{n}|\partial _{\lambda }H|\phi _{m}\rangle $%
. To see the validity of the adiabatic theorem, we need to address the fidelity
between $|\phi _{0}(t)\rangle $ and $|\Psi (t)\rangle $. For the normalized
states $|\phi _{0}(t)\rangle $ and $|\Psi (t)\rangle$, the fidelity is
\begin{equation}
F=|\langle \phi _{0}(t)|\Psi (t)\rangle|.  \label{eq:fidelity332}
\end{equation}
The adiabatic theorem requires that $F\simeq 1$.

However, it is not easy to estimate exactly the values of the integrals in Eqs.
(\ref{eq:berrycon}) and (\ref{eq:hopingcon}). To see the qualitative behavior
of the leading term of the fidelity, we first make use of $|\phi _{n}(\lambda
)\rangle $ as reference states, then the energy levels vary slowly with time.
Under this approximation, the fidelity is the same as the perturbative form of
the Loschmidt echo \cite{JZhang}
\begin{equation}
F_{1}\simeq 1-\left( \delta \lambda \right) ^{2}\sum_{n\neq 0}\frac{%
|H_{I}^{n0}|^{2}[1-\cos \left( \omega _{0n}\Delta t\right) ]}{\omega
_{0n}^{2}}.  \label{eq:fidelityup}
\end{equation}%
Here the second term denotes the transition probability and a phase factor from
Eq. (\ref{eq:berrycon}) has been normalized out.

The second alternative approach is to find the bound of the integral in Eq. (%
\ref{eq:hopingcon}). Because of%
\begin{equation}
\left\vert \exp \left( -i\int^{t}\omega _{0m}dt^{\prime }\right) \right\vert
=1,
\end{equation}%
we have
\begin{equation}
|\tilde{a}_{m}|\leq \frac{1}{\tau _{0}}\int_{0}^{\Delta t}\left\vert \frac{%
H_{I}^{m0}}{\omega _{0m}}\right\vert dt.
\end{equation}%
Then we obtain a lower bound of the fidelity
\begin{equation}
F_{2}\simeq 1-\frac{\left( \delta \lambda \right) ^{2}}{2}\sum_{n\neq 0}%
\frac{|H_{I}^{n0}|^{2}}{\omega _{0n}^{2}},  \label{eq:fidelitylower}
\end{equation}%
hence an upper bound of the transition probability.

Mathematically, $F_1$ defines a distance between $|\phi_0(\lambda)\rangle $ and
$|\Psi (t)\rangle$ and $F_2$ a distance between $|\phi_0(\lambda)\rangle $ and
$|\phi_0(\lambda+\delta\lambda)\rangle$. Therefore, $F_1$, $F_2$, and $F$ in
Eq. (\ref{eq:fidelity332}) for a ``triangle" in the parameter space. Our
concern here is that the transition probabilities in both $F_1$ and $F_2$ are
determined by the fidelity susceptibility \cite{WLYou07,PZanardi0701061}
\begin{equation}
\tilde{\chi}_{F}=\sum_{n\neq 0}\frac{|H_{I}^{n0}|^{2}}{\omega _{0n}^{2}},
\label{eq:fsfsfsg}
\end{equation}%
which defines also the scale of the original fidelity $F$ defined in Eq.
(\ref{eq:fidelity332}). In previous studies on the quantum adiabatic theorem,
the formulism given in Eqs. (\ref{eq:berrycon}-\ref{eq:fidelitylower}) are
familiar to us, however, few attention has been paid to the scaling behavior of
the quantity (the fidelity susceptibility) until recently
\cite{LCVenuti07,SJGu072}.

For a $d$-dimensional system, the fidelity susceptibility of the driving
Hamiltonian has its own dimension $d_{a}$ \cite{SJGu08} instead of the
system's real dimension though in many cases both dimensions are equal. That
is
\begin{equation}
\tilde{\chi}_{F}\propto L^{d_{a}}
\end{equation}%
given that $L$ is larger enough. In the critical region, the quantum adiabatic
dimension $d_{a}=2d+2\zeta -2\Delta _{V}$\cite{LCVenuti07} with $d,\zeta $, and
$\Delta _{V}$ being the real dimension, dynamic exponent and scaling dimension
of the driving Hamiltonian respectively. Clearly, in this case, the quantum
adiabatic dimension $d_{a}$ can be larger than $d$. In the non-critical region,
the correlation length is finite, then we usually have $d_{a}=d$ or $d_{a}<d$.
For instance, in the fully polarized phase of the LMG model, $d_{a}=0$
\cite{HMKwok07} (here the LMG model is considered as a one-dimensional system
with infinite-range interactions). In Table \ref{tab:critcalexp}, we show the
adiabatic dimension for three exactly solvable models, i.e. the one-dimensional
transverse-field Ising model \cite{Sachdev}, the LMG model \cite{LMGmodel}, and
the Kitaev honeycomb model \cite{Kitaev}, around their corresponding phase
transition point. These data are collected from the recent fidelity approaches
to QPTs, as shown in the caption of the table.

\begin{table}[tbp]
\caption{The adiabatic dimensions $d_{a}^{+}, d_{a}^{c}$, and $d_{a}^{-}$ (
above, at, and below the critical point) for the 1D Ising model \protect\cite%
{Zanardi06}, the LMG model \protect\cite{HMKwok07}, and the Kitaev honeycomb
model (KHM) \protect\cite{SYang08FS,SJGu08}.}
\label{tab:critcalexp}
\begin{center}
\begin{ruledtabular}
\begin{tabular}{c c c c c}
 Model (critical point) & $d$ & $d_a^c $ & $d_a^+ $  & $d_a^-$  \\
\hline 1D Ising model ($h_c=1$) & 1 & 2  & 1 & 1
\\
\hline LMG model($h_c=1$) & 1 & 4/3 & 0
& 1  \\
\hline KHM($J_c=1/2$) & 2 & 5/2  & 2
& 2+ln  \\
\end{tabular}
\end{ruledtabular}
\end{center}
\end{table}

However, the results obtained from the perturbation theory is valid only if the
change in the driving Hamiltonian is very small. Physicists are interested in
the case that the change in the Hamiltonian, though varies slowly with the
time, is not small, then the perturbation method can not be applied directly.
To solve this problem, we first assume, without loss of generality, that the
system evolves from $\lambda _{i}$ to $\lambda_{f}$ during the time interval
$\tau _{0}$, and the quantum adiabatic dimension in this region is $d_{a}$.
Secondly, we divide the interval $\lambda _{f}-\lambda _{i}$ into $M$
subintervals (Here $M$ defines also the scale of $\tau_0$), that is $ \delta
\lambda ={(\lambda _{f}-\lambda _{i})}/{M}$ . The total leading transition
probability to excited states $P_{t}$ then scales like
\begin{equation*}
P_{t}\sim M\left( \frac{1}{M}\right) ^{2}\tilde{\chi}_{F}.
\end{equation*}%
Because of $\tilde{\chi}_{F}\propto L^{d_{a}}$, $M$ should be \emph{at least}
about the order of $L^{d_{a}}$ to ensure the validity of the perturbation
method. The final (minimum) probability of staying in the ground state $|\phi
_{0}\rangle $ at $\lambda _{f}$ then becomes
\begin{equation}
P_{s}\simeq \left[ 1-\frac{1}{2}\left( \frac{\Delta t}{\tau _{0}}\right) ^{2}%
\tilde{\chi}_{F}\right] ^{L^{d_{a}}}.
\end{equation}%
The quantum adiabatic theorem requires $P_{s}\simeq 1$. In the thermodynamic
limit, we arrive at a simple inequality,
\begin{equation}
\tau _{0}\gg \kappa L^{d_{a}},  \label{eq:durationtime}
\end{equation}%
where $\kappa $ is a $L$-independent quantity. The above inequality
concludes the key result of the present work.

Clearly, the inequality (\ref{eq:durationtime}) defines the scale of duration
time required by the quantum adiabatic theorem. That is, a sufficient adiabatic
condition should defines the duration time in order of $L^{d_a}$. For a
realistic (thermodynamic) system, the only comparable scale is the system size
$N=L^{d}$, which is about the order of the Avogadro constant ($6.02\times
10^{23}$). If we require that a physically acceptable duration time is, at
most, proportional to the system size, then the two limits of $%
N\longrightarrow \infty $ and $\tau _{0}\longrightarrow \infty $ do not commute
with each other if $d_{a}>d$, and the quantum adiabatic theorem might be
violated. On the other hand, in case that the starting and ending points are
located in two regions with different quantum adiabatic dimensions, say
$d_{a1}$ and $d_{a2}$, the required duration time is determined by $\tau_{0}\gg
\kappa L^{{\max }(d_{a1},d_{a2})}$.

Now we take the one-dimensional transverse-field Ising model as an example to
illustrate the validity of the quantum adiabatic theorem according to our
criterion. The Hamiltonian of the Ising model reads
\begin{eqnarray}
H &=&-\sum_{j=1}^{N}\left( \sigma _{j}^{x}\sigma _{j+1}^{x}+h\sigma
_{j}^{z}\right) ,  \label{eq:Hamiltonian_Ising0} \\
\sigma _{1}^{x} &=&\sigma _{N+1}^{x},
\end{eqnarray}%
where $h(t<0)=-t/\tau _{0}$. The Hamiltonian (\ref{eq:Hamiltonian_Ising0})
has been used as a prototype model in both studies on the ground-state
fidelity \cite{Zanardi06} and dynamics of QPTs \cite{Zurek,Dziarmaga}. If
the quantum adiabatic theorem holds true, $t<-\tau _{0}$ corresponds to the
paramagnetic phase, and $-\tau _{0}<t<0$ is the ferromagnetic phase. A
second order QPT occurs at $t=-\tau _{0}$. The fidelity susceptibility in
the whole region can be calculated as
\begin{equation}
\tilde{\chi}_{F}=\sum_{k>0}\left( \frac{d\theta _{k}}{dh}\right) ^{2},
\label{eq:fsisingana}
\end{equation}%
with $k=\pi /N,3\pi /N,\cdots ,\pi (N-1)/N$, and%
\begin{equation}
\frac{d\theta _{k}}{dh}=\frac{1}{2}\frac{\sin k}{1+h^{2}-2h\cos k}.
\end{equation}%
It can be shown that $\tilde{\chi}_{F}\varpropto N^{2}$ for $h=1$, while $%
\tilde{\chi}_{F}\varpropto N$ for $h\neq 1$. Therefore, according to our
criterion, if the starting and ending point are in the same phase, i.e. $%
t_{i}(t_{f})<-\tau _{0}$ or $-\tau _{0}<t_{i}(t_{f})<0$, the duration time
required by the adiabatic condition is $N\ll \tau _{0}$. However, if $%
t_{i}<-\tau _{0}$ and $-\tau _{0}<t_{f}<0$, the system will across the
transition point $h=1$, at which the adiabatic dimension is 2. So the required
duration time should satisfy $\tau_{0}\gg \kappa N^{2}$. This observation
is consistent with result obtained via the Landau-Zener formula \cite%
{LDLandau,CZener32} in the recent studies on quench dynamics in the Ising
model \cite{Dziarmaga}. Therefore, the quantum adiabatic theorem might break
down at the critical point.

\begin{figure}[tbp]
\includegraphics[width=8cm]{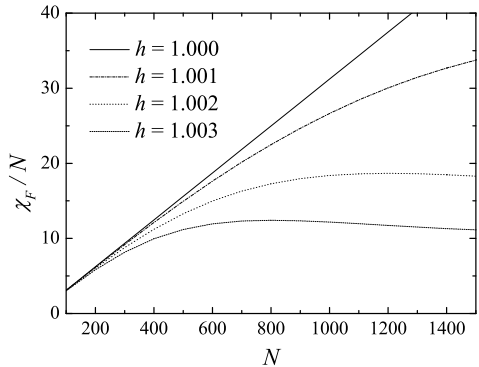}
\caption{The scaling behavior of the fidelity susceptibility around the
critical point of the one-dimensional transverse-field Ising model.}
\label{figure.eps}
\end{figure}

However, the problem is still subtle because the transition point is not a
region but a \textquotedblleft point". Then the large $N$ behavior might be
quite different from that of the infinite limit. In Fig. \ref{figure.eps},
we show the scaling behavior of the fidelity susceptibility around the
critical point. We can see that only at the critical point, $\tilde{\chi}%
_{F}/N\propto N$. While away from the critical point, though the closer to
the critical point, the larger the fidelity susceptibility, the later will
be finally saturated to
\begin{equation}
\frac{\chi _{F}}{N}=\left\{
\begin{tabular}{ll}
$\frac{1}{16(1-h^{2})}$ & for $h<1$ \\
$\frac{1}{16h^{2}(h^{2}-1)}$ & for $h>1$%
\end{tabular}%
\right. ,
\end{equation}%
respectively as $N$ increases. Therefore, for any simulation on a large but
finite sample, the duration time should satisfy $\tau _{0}\gg \kappa N^{2}$ in
the region close enough to the critical point. While in the infinite $N$ limit,
the condition $\tau _{0}\gg\kappa N^{2}$ is valid rigorously only at the
critical point.

Moreover, we can see from Table I that the quantum adiabatic theorem does not
hold true at the critical point of both the LMG model and Kitaev honeycomb
model also. On the other hand, it has been found recently that the quantum
adiabatic dimension in the gapless phase of the Kitaev honeycomb model is 2+ln
\cite{SJGu08}, which is larger than the real dimension 2. Our quantum adiabatic
condition implies that the quantum adiabatic theorem might be violated in the
whole gapless phase of the Kitaev honeycomb model.

In summary, we have proposed the quantum adiabatic condition for quantum
systems in the thermodynamic limit. A general inequality between duration time
required by the quantum adiabatic theorem, the system size, and quantum
adiabatic dimension are established. For the commonly studied linear quenches,
our results show that the adiabatic condition might be violated if the
adiabatic dimension is larger than the real dimension. This phenomenon usually
occurs at the quantum critical point and those strange phases of $d_a>d$.

We thank L. G. Wang, Z. G. Wang, and Yi Zhou for helpful discussions and
comments. This work is supported by the Earmarked Grant Research from the
Research Grants Council of HKSAR, China (Project No. CUHK 400807).

\end{document}